\documentclass[useAMS,usenatbib]{mn2e}
\usepackage{graphicx}
\usepackage{amssymb}

\newcommand{\placetabone}{
\begin{table}
    \centering
    \renewcommand{\footnoterule}{}  
    \begin{tabular}{ccc}
      \hline \hline
      \textbf{BH} & \textbf{Collision} & \textbf{BH softening}\\
      \textbf{Mass ratio}   & \textbf{angle} ($^{\rm o}$) & (pc)\\
      \hline
	1:1 & 0 & 0.5\\
	1:1 & 30 & 0.5 \\
	1:1 & 60 & 0.5 \\
	1:1 & 90 & 0.5\\
	1:1 & 120 &0.5\\
	1:1 & 150 &0.5\\
	1:1 & 180 & 0.5\\
	\hline
	1:5 & 0 & 0.5\\
	1:5 & 30 & 0.5\\
	1:5 & 60 & 0.5\\
	1:5 & 90 & 0.5\\
	1:5 & 120 &0.5\\
	1:5 & 150 &0.5\\
	1:5 & 180 & 0.5\\
	\hline
	1:10 & 0 & 0.5\\
	1:10 & 30 & 0.5\\
	1:10 & 60 & 0.5\\
	1:10 & 90 & 0.5\\
	1:10 & 120 &0.5\\
	1:10 & 150 &0.5\\
	1:10 & 180 & 0.5\\
      \hline
    \end{tabular}
    \caption[Set of simulations]{Set of simulations with mass ratios,
      collision angles and the gravitational softening of the
      secondary SMBH. Collision angles $i_{\rm BH}$ from 0 to 60$^{\rm
        \circ}$ are prograde encounters whereas $i_{\rm BH}$ values
      larger than 90$^{\rm \circ}$ denote retrograde ones.}
    \label{tab:sims}
\end{table}
}

\newcommand{\placetabtwo}{
\begin{table}
    \centering
    \renewcommand{\footnoterule}{}  
    \begin{tabular}{ccccc}
      \hline \hline
      {} & \multicolumn{2}{c}{\textbf{$f_{phot}$}} & \multicolumn{2}{c}{\textbf{$f_{kin}$}}\\
      \hline
      \textbf{Mass ratio} & \textbf{raw} & \textbf{sin($i_{\rm BH}$)} & \textbf{raw} & \textbf{sin($i_{\rm BH}$)}\\
      \hline
	1:1    &    19.0    &   16.1    &    38.1    &    45.5\\    
	1:5    &    61.9    &   72.8    &    71.4    &    80.1\\    
 	1:10   &    85.7    &   85.0    &   100.0    &   100.0\\    
      \hline
    \end{tabular}
    \caption[Disc detection fractions]{Fractions in percent of merger
      remnants where the disc can still be detected, for the three
      mass ratios and according to whether the disc is detected
      photometrically or kinematically. The ``raw'' columns list
      uncorrected fractions, whereas the ``sin($i_{\rm BH})$'' columns
      list fraction that are corrected by this factor to take into
      account that co-planar encounters are much less likely to happen
      than those with higher $i_{\rm BH}$ inclinations.}
    \label{tab:frac}
\end{table}
}

\newcommand{\placefigone}{
\begin{figure}
\begin{center}
\includegraphics[width=\columnwidth]{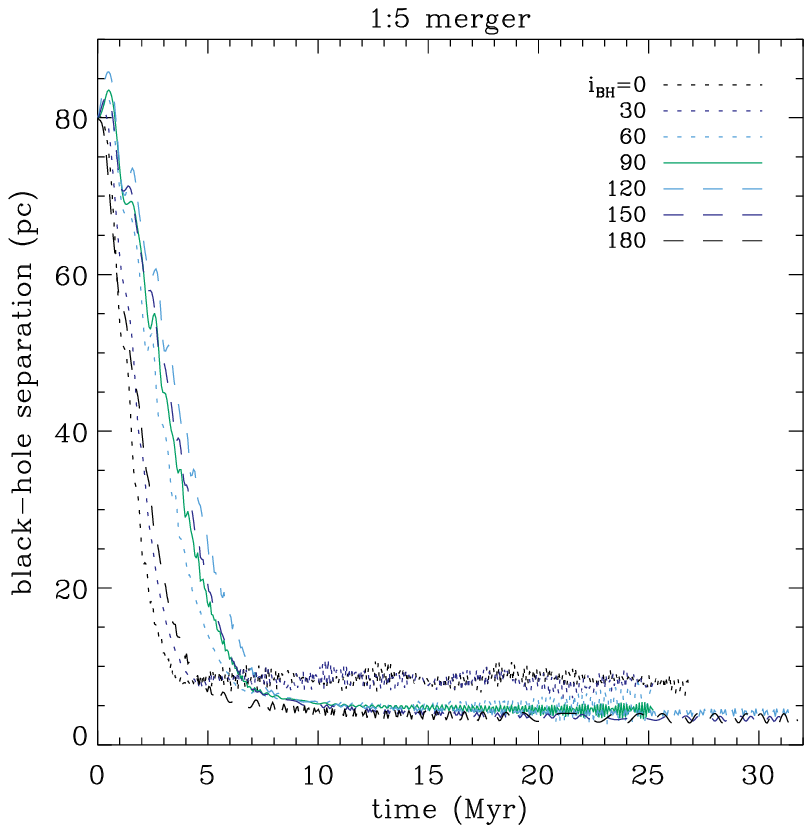}
\end{center}
\caption[1:5 BH separations]{Time evolution for the separation between
  the primary and secondary black holes for the 1:5 mass ratio
  encounters. Dotted lines denote prograde encounters, whereas dashed
  lines represent retrograde ones. The case of 1:1 (1:10) mergers
  would follow similar decay patterns although at shorter (longer)
  time scales.}
\label{fig:sep_1to5}
\end{figure}
}

\newcommand{\placefigtwo}{
\begin{figure}
\begin{center}
\includegraphics[width=\columnwidth]{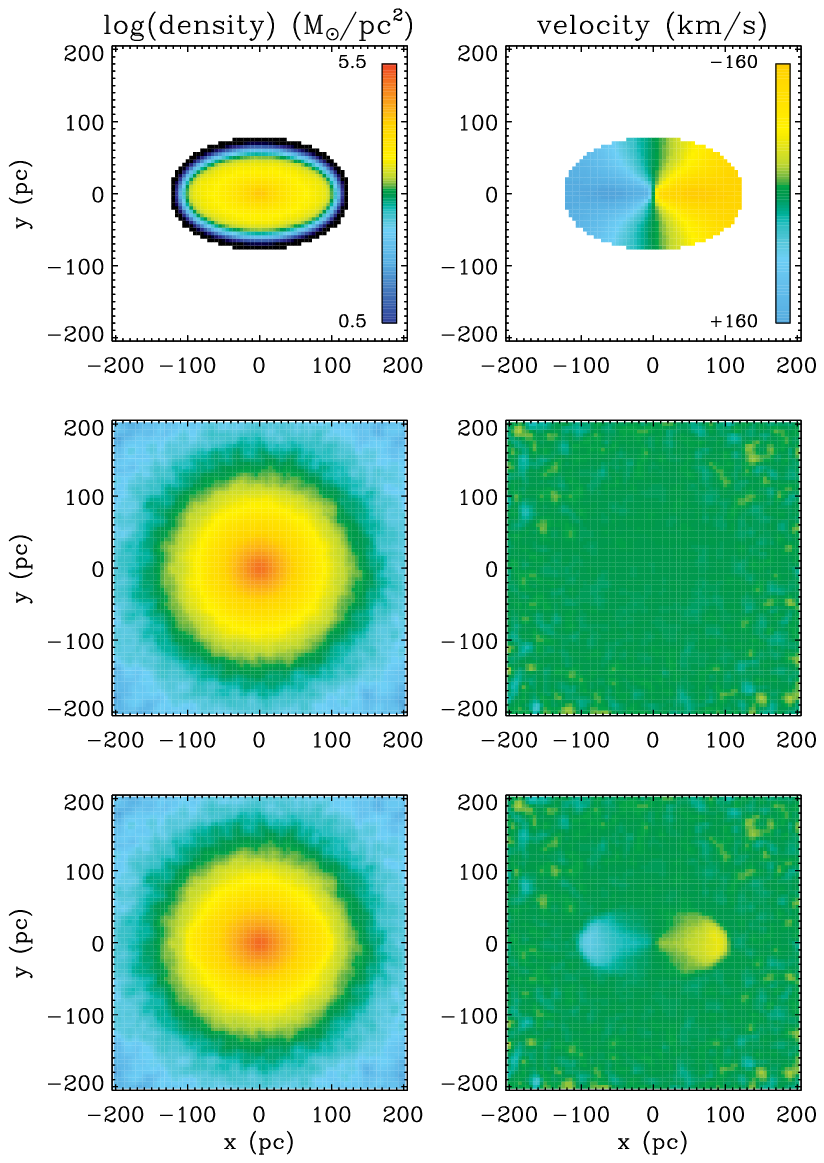}
\end{center}
\caption[Simulation initial conditions]{Surface mass density (left
  panels, in $\rm M_{\odot}\;pc^{-2}$ and in logarithmic scale) and
  velocity fields (right panels in $\rm km\;s^{-1}$) for the initial
  conditions of our simulations, showing, from top to bottom the disc
  particles, the bulge particles and the entire system. This figure
  zooms in a 400 x 400 pc region around the centre that encompasses
  the entire disc, which is being viewed at a 60$^{\rm \circ}$
  inclination.}
\label{fig:all_i}
\end{figure}
}

\newcommand{\placefigthree}{
\begin{figure}
\begin{center}
\includegraphics[width=\columnwidth]{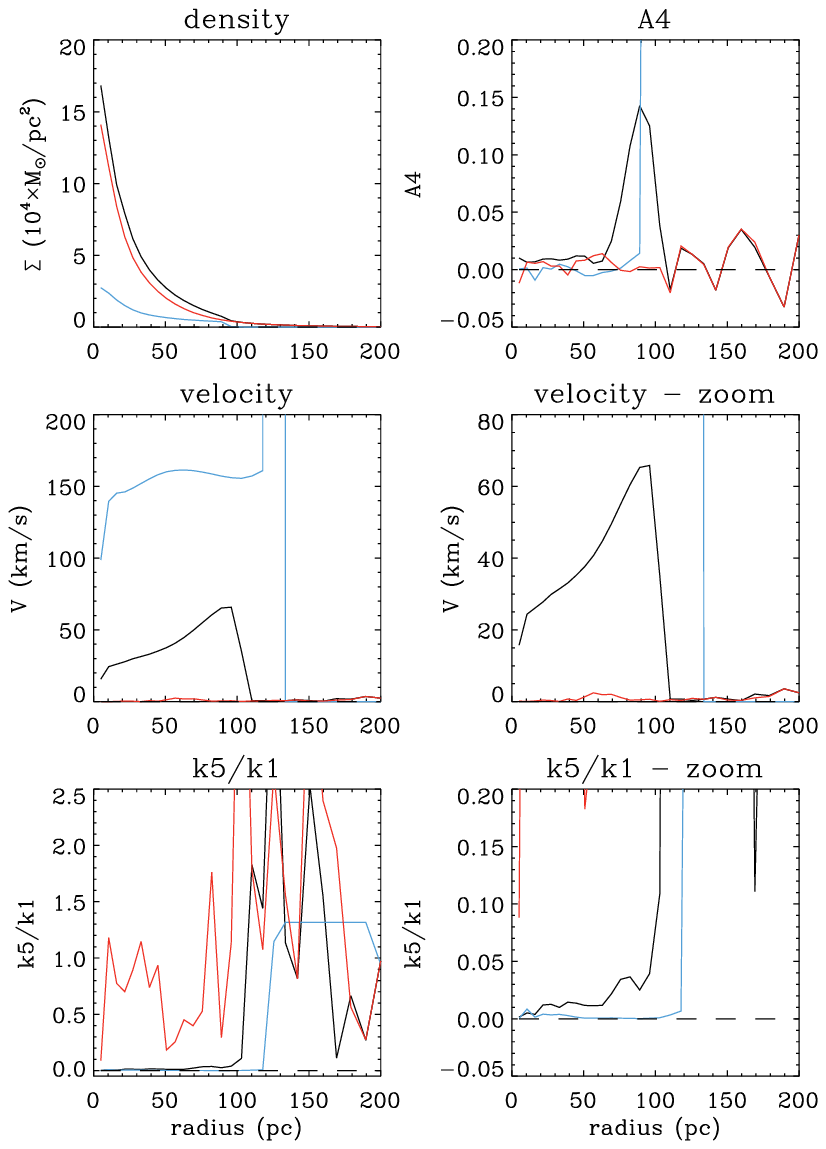}
\end{center}
\caption{\textsc{Kinemetry} results for the initial conditions shown in
  Fig.\ref{fig:all_i}. The blue, red and black lines show radial
  trends for the surface brightness, $a_{4}$, velocity and $k5/k1$ of
  the disc, bulge and all particles, respectively and as indicate at
  the top of each panel. A peak in the $a_{4}$ and low values of
  $k5/k1$ indicate the presence of a disc.}
\label{fig:d_f}
\end{figure}
}

\newcommand{\placefigfour}{
\begin{figure*}
\begin{center}
\includegraphics[width=\textwidth]{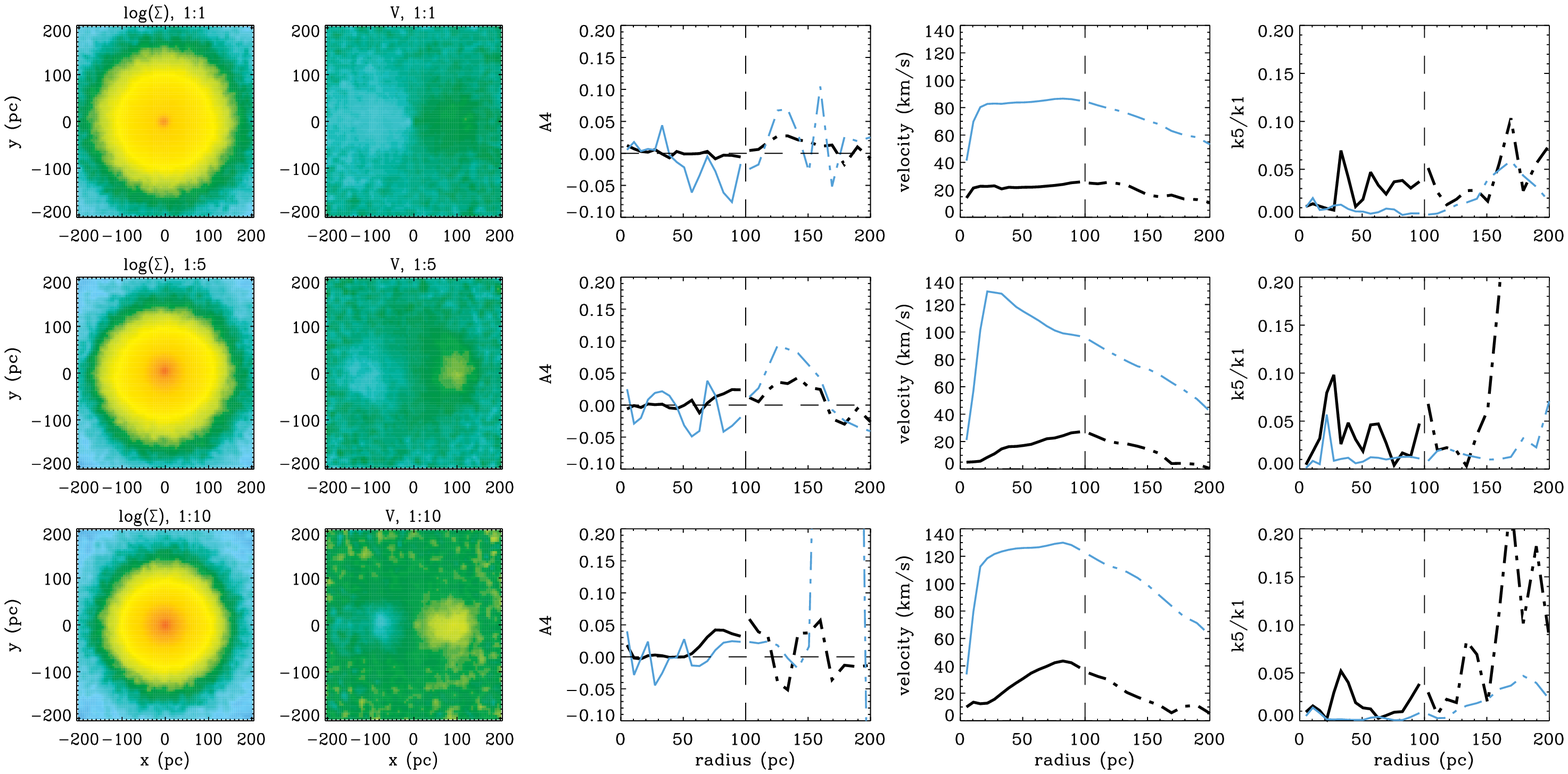}
\end{center}
\caption{Final total surface density maps and velocity fields (left
  panels) and correspoding kinemetric profiles for the $a_4$
  parameter, the velocity $V$ and the $k5/k1$ ratio (central and right
  panels) resulting from 1:1, 1:5 and 1:10 mass ratio encounters (top,
  middle and lower rows) with $i_{\rm{BH}} = 30^\circ$ and as viewed
  from an inclination $i = 60^\circ$. In the panels showing our
  \textsc{kinemetry} results, the black lines show these measurements for the
  entire system whereas the blue lines are for the disc particles
  only. Vertical dashed lines indicate the original extent of the
  disc. We compute average values for those \textsc{kinemetry} parameters
  inside this radius of 100 pc, except for the most edge-on
  projections in the case of $V$ and $k5/k1$, where we restricted our
  analysis to the inner 50 pc.}
\label{fig:kinemetry}
\end{figure*}
}

\newcommand{\placefigfive}{
\begin{figure*}
\begin{center}
\includegraphics[width=\textwidth, bb = 30 30 520 605]{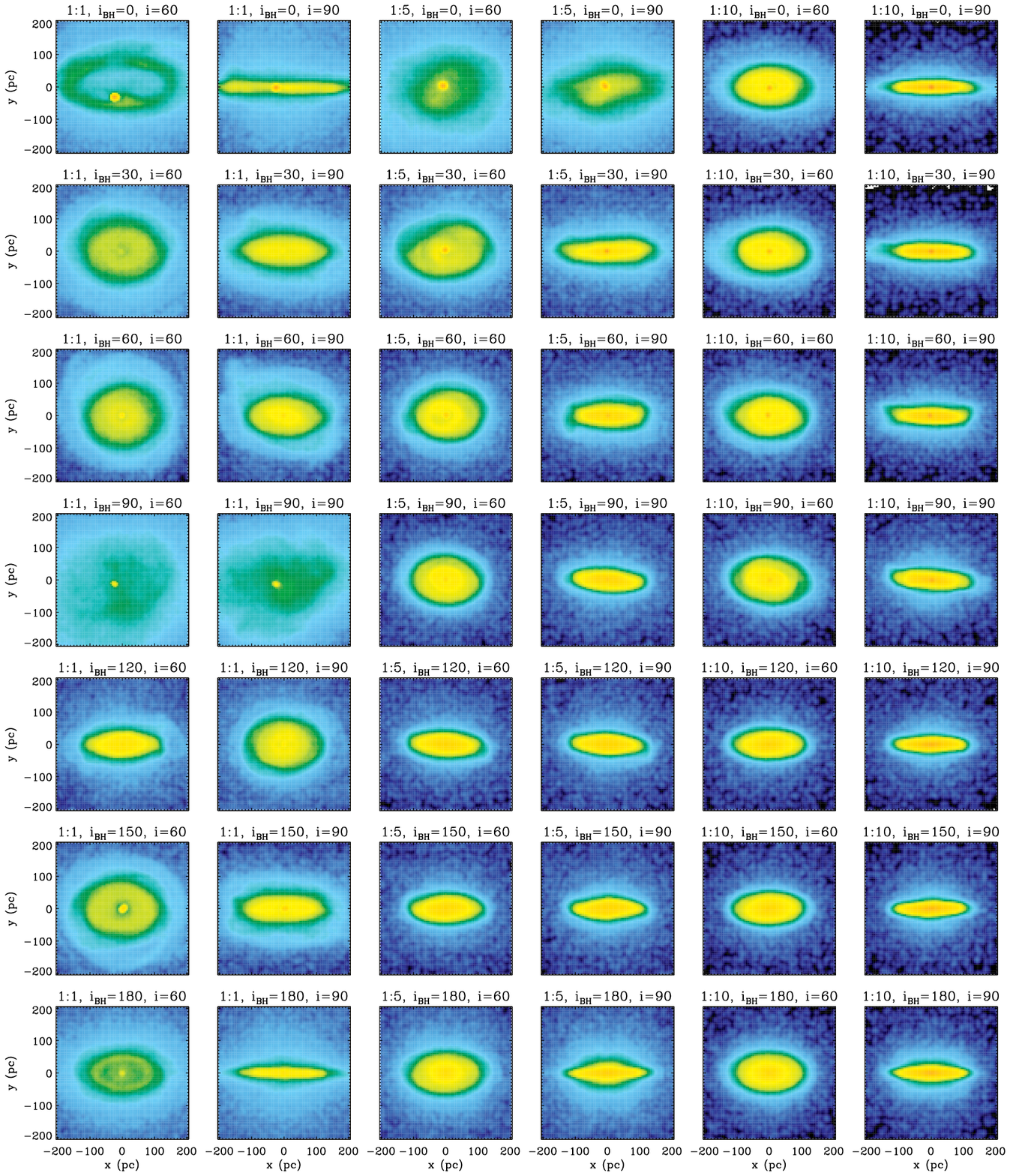}
\end{center}
\caption{Surface density for the disc particles at the end of our
  simulations. From left to right we show the disc remnants of 1:1,
  1:5 and 1:10 mergers, both from an viewing angle of 60$^{\circ}$ and
  as seen edge-on, whereas from top to bottom we display simulations
  with increasing impact $i_{\rm BH}$ angles. The maps adopt the same
  logarithmic scaling and units as shown in Fig.~\ref{fig:all_i}.}
\label{fig:discmaps}
\end{figure*}
}

\newcommand{\placefigsix}{
\begin{figure*}
\begin{center}
\includegraphics[width=\textwidth]{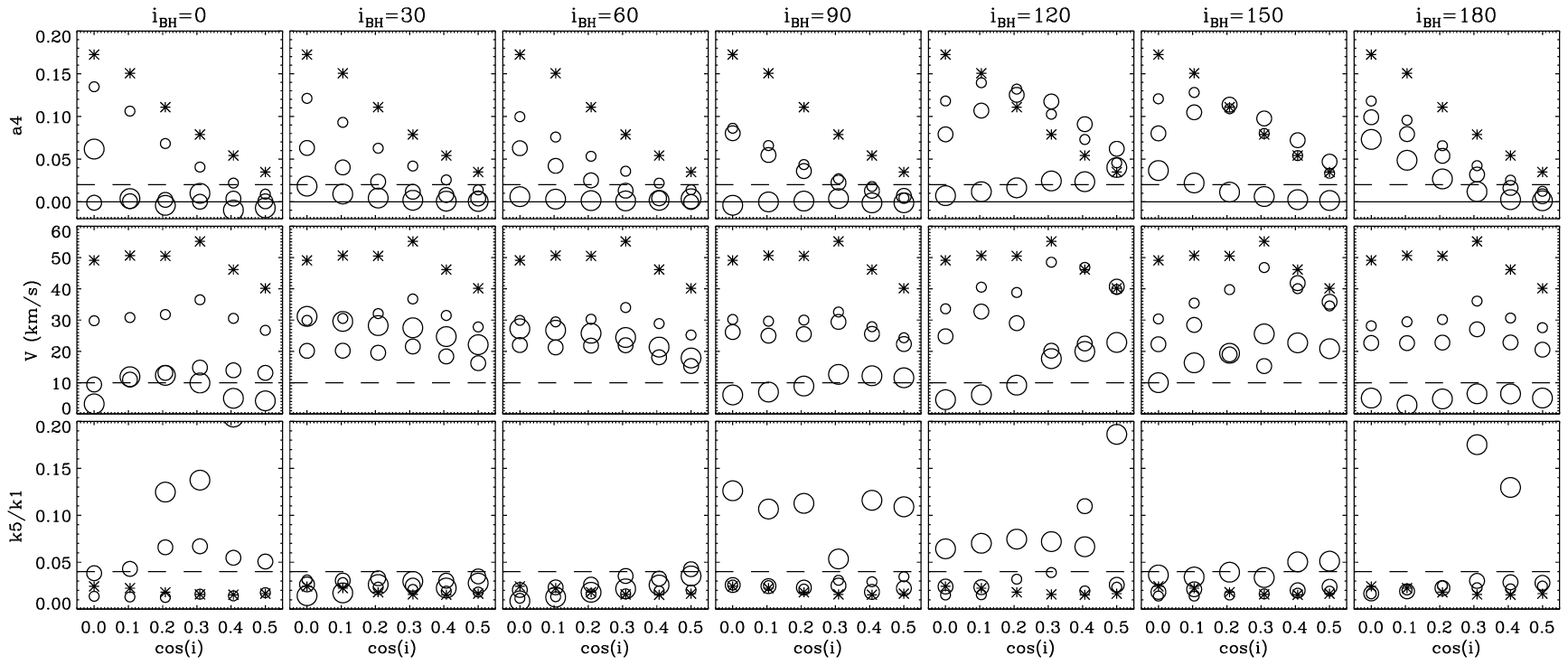}
\end{center}
\caption[]{Average values for the $a_{4}$ parameter (top row),
  velocity $V$ (middle row) and the $k5/k1$ ratio (lower row) as
  measured at the final step of each of our simulations and as a
  function of our viewing angle ($\cos{i}$). The different panels from
  left to right show the results from simulations of similar values
  for the collision angle $i_{\rm BH}$, with the size of the circles
  in each panel tracing the average $a_{4}$, $V$ and $k5/k1$ values
  for simulations with increasing merger mass ratio (i.e., from 1:10,
  1:5 to 1:1). As a reference, in each panel the asterisks indicate
  the corresponding values as measured in our initial conditions,
  whereas the dashed horizontal line indicates our threshold for
  detecting the photometric or kinematic signatures of a disc in the
  merger remnants (i.e., discy isophotes, coherent rotation and
  disc-like kinematics in the case as indicated by the $a_{4}$, $V$
  above and $k5/k1$ below the thresholds, respectively).}
\label{fig:kinresults}
\end{figure*}
}

\title[Nuclear stellar disc fragility]{On the fragility of nuclear
  stellar discs against galaxy mergers: surviving photometric and
  kinematic signatures of nuclear discs} \author[Sarzi, Ledo \&
  Dotti]{M. Sarzi$^{1}$\thanks{E-mail: m.sarzi@herts.ac.uk},
  H. R. Ledo$^{1}$, M. Dotti$^{2,3}$\\
   $^{1}$Centre for Astrophysics Research, University
   of Hertfordshire, College Lane, Hatfield AL10 9AB, United
   Kingdom\\ 
   $^{2}$Dipartimento di Fisica G. Occhialini, Universit\`a
   degli Studi di Milano, Bicocca, Piazza della Scienza 3, 20126
   Milano, Italy\\ 
   $^{3}$INFN, Sezione di Milano-Bicocca, Piazza della
   Scienza 3, 20126 Milano, Italy}

\begin{document}


\pagerange{\pageref{firstpage}--\pageref{lastpage}} \pubyear{2015}

\maketitle

\label{firstpage}

\begin{abstract}
Nuclear stellar discs (NSDs) can help to constrain the assembly
history of their host galaxies, as long as we can assume them to be
fragile structures that are disrupted during merger events. In this
work we investigate the fragility of NSDs by means of N-body
simulations reproducing the last phases of a galaxy encounter, when
the nuclear regions of the two galaxies merge. For this, we exposed a
NSD set in the gravitational potential of the bulge and supermassive
black hole of a primary galaxy to the impact of the supermassive black
hole from a secondary galaxy. We explored merger events of different
mass ratios, from major mergers with a 1:1 mass ratio to intermediate
and minor interactions with 1:5 and 1:10 ratios, while considering
various impact geometries.
We analyse the end results of such mergers from different viewing
angles and looked for possible photometric and kinematic signatures of
the presence of a disc in the remnant surface density and velocity
maps, while adopting detection limits from real observations.
Our simulations show that indeed NSDs are fragile against major
mergers, which leave little trace of NSDs both in images and velocity
maps, while signatures of a disc can be found in the majority of the
intermediate to minor-merger remnants and in particular when looking
at their kinematics.
These results show that NSDs could allow to distinguish between these
two modes of galaxy assembly, which may indeed pertain to different
kinds of galaxies or galactic environments.
\end{abstract}

\begin{keywords}
galaxies: elliptical and lenticular, cD -- galaxies: evolution,
galaxies: nuclei, galaxies: interactions, galaxies: kinematics and
dynamics

\end{keywords}

\section[]{Introduction}

Nuclear stellar discs (NSDs) were first discovered in images taken
with the Hubble Space Telescope \citep[HST, e.g.,][]{Jaf94, vdB94},
which allows for the detection of such structures that are often of
only tens to hundreds of parsecs across and which reside mainly in the
centre of early-type galaxies \citep{Piz02}.  Previous studies have
focused on the properties of few of such discs \citetext{e.g.,
  \citealp{Mor04}, \citealp{Piz02}, \citealp{Sco98}} but now thanks to
the census of \citet[][hereafter Paper I]{Led10} we have a
sufficiently large sample to study these features more systematically.

NSDs are not only interesting tools for determining the masses of
supermassive black holes \citep[SMBH,][]{vdB96}, but can also be a
powerful tool to trace galaxy merging history. In Paper I we have
shown evidence pointing to the destruction of these discs during a
merger event, even though if there is gas involved during such an
event, part of this gas can fall to the centre and form a new stellar
disc.  Therefore, when present, NSDs provide us with a look-back time
to the last merger experienced by their host galaxies, a precise one
if they were born at that moment, or at least a lower limit if they
formed afterward. This look-back time can be used to test and improve
galaxy formation scenarios and the predictions of semi-analytical
models \citep[e.g.,][]{Kho06a, DeL06}. In fact, NSDs are particularly
apt for this role since their stellar age can be determined very
precisely. Typical degeneracies between the stellar age and
metallicity that affect population studies can indeed be reduced using
integral-field spectroscopic data and by knowing in advance
\citep[thanks to a photometric disc-bulge decomposition][]{Sco95} the
disc-light contribution (Sarzi et al., in preparation).  What is still
unknown to us is understanding more precisely to which extent these
discs are fragile and what kind of merging events they actually
trace. Do they trace only major mergers (as shown in the preliminary
simulations published in Paper I), or do smaller satellite galaxies
also have the ability to destroy them? If so, how small and in which
circumstances?

The impact of mergers has previously been explored in the case of
kinematically decoupled cores \citep[KDC, see][]{Boi11} or metallicity
gradients \citep{Kob04}, but so far little is known about the survival
of NSDs.
In fact, in this study we are primarily interested in establishing
whether the presence of an NSD would still be detectable in the merger
remnant by looking for the typical photometric and kinematic
signatures of NSDs, such as discy isophotes and regular disc-like
rotation, as observers would do on real galaxy images or velocity
fields. At the same time, we will resort to our view of the disc
particles at the end of our simulations to better understand the disc
disrupting process as a function of the parameters that we will
explore, checking also for false positive detections.

This work is organised as follows. In Sec.~\ref{sec:simsetup} we
describe the set up of our initial conditions and the various kinds of
merging simulations that we considered. In Sec.~\ref{sec:simanalysis}
we describe the construction of surface density and velocity maps for
our merger remnants as seen from different viewing angles and our use
of the technique called \textsc{kinemetry} \citep{Kra06} to analyse these maps
and look for both photometric and kinematic signature of discs. In
Sec.~\ref{sec:results} we then assess how often such signatures would
indicate the presence of a disc, and finally discuss our findings in
Sec.~\ref{sec:conclusions}.

\section{Simulation thought and design}
\label{sec:simsetup}

We are interested in understanding how a galaxy merger would affect
the innermost parts of a galaxy with a small nuclear disc embedded at
its center.
This is likely to depend on the nature of the encounter, for instance
on it being either a major or minor merger, on the geometry of the
collision, such as prograde as opposed to retrograde, and finally on
the mass of the central disc, given that more massive discs should be
more stable. In this paper we deal only with the magnitude and
geometry of the perturbing event, and focus on nuclear stellar discs,
which can extend up to a few 100 pc (Paper~I) and can have stellar
masses comparable to that of their host galaxy SMBH \citep{Sco98}.
Furthermore, we conducted a series of simulations meant to look only
at the final stages of a merging event, when the SMBH hole of the
perturbing galaxy reach the central regions of the galaxy hosting the
nuclear disc and a primary SMBH.
Dynamical friction will indeed cause such a secondary black hole to
sink towards the central regions of the primary galaxy or what would
eventually become the merger remnant. Most of the stellar cusp around
the secondary is likely to be stripped during the early stages of the
galaxy pairing, even though the total stellar mass that would remain
bound to the secondary could depend on a plethora of factors, such as
the initial mass concentration of the secondary galaxy or the mass
ratio between the merging galaxies, the orbital parameters of the
mergers, and so on \citep[][]{Cal09,VWa14}. 
At present we therefore decided to start by considering the simplest
case of a naked secondary black hole.

\placefigone
\placetabone

After setting stable initial conditions for a $10^8 \rm{M}_{\odot}$
nuclear disc with a radius of 100 pc that is embedded in a bulge with
its own $10^8 \rm{M}_{\odot}$ SMBH, we introduce in
the simulations an additional massive particle at an initial distance
of 80 pc from the centre, on a circular orbit. This particle mimics
the presence of a perturbing SMBH (the secondary, hereafter) as it is
reaching the centre of the primary galaxy during the last stages of a
galaxy merger. Inevitably, this secondary SMBH interacts with the star
particles of the primary nucleus and sinks toward the centre while
perturbing the distribution of the stars. After the secondary had
reached the centre, the simulations were let to evolve for several
dynamical times as to allow the system to relax to a point where we
would not expect further perturbations to the dynamics of the
stars. It is at this point that we will proceed to analyse the merger
remnant, looking in particular to detect the same photometric and
kinematics signatures that observers would regard as indicative of the
presence of a nuclear stellar disc.

The simulations were conducted using GADGET-2 \citep{Spr05}, a
parallel code designed for cosmological N-body/smoothed particle
hydrodynamics simulations. Each simulation involved 600K stellar
particles and two additional collisionless particles to model two
SMBHs.
The spatial resolution of our simulations is determined by the
gravitational softening of the particles, which is equal to 0.5 pc for
all the particles. 300K particles have been used to model the nuclear
stellar disc of total mass $M_{\rm disc}=10^8 \rm{M}_{\odot}$ and an
initial outer radius of 100 pc. The disc surface density, $\Sigma$,
follows a Mestel \citep{Mes63} profile where $\Sigma \propto 1/R$, and
has an aspect ratio $H/R \approx 0.05$, similarly to previously used
nuclear discs \citep[e.g.,][]{Dot07,Dot06,Esc05} and to the nuclear
discs outcome of cosmological simulations \citep[e.g.,][]{Lev08}.
Another 300k particles have been used to model the central regions of
the spherical stellar bulge hosting the disc. The bulge follows a
Plummer profile
\begin{equation}
 \rho ={3 \over 4 \pi}{M_{\rm bulge}\over b^3} \left(1+{r^2\over
   b^2}\right)^{-5/2}
\end{equation}
where $b$ is the core radius ($b = 50\rm pc$), $r$ the radial
coordinate and $M_{\rm bulge}$ the total mass of the spheroid, with
$M_{\rm bulge} = 6.98 M_{\rm disc}$.  Finally, the bulge and disc
stellar structure hosts a $10^8 \mathrm{M}_{\odot}$ SMBH. While the
SMBH is initially at rest at the centre of the structure, the
velocities of the star particles modelling the disc and the bulge are
chosen to set the system in dynamical equilibrium. Since an analytic
formulation of the distribution function of the composite structure is
not available, we let the system to stabilise into dynamical
equilibrium in isolation. We further verified that over the typical
timescale of our simulated encounters (a few ten Myr, see
Fig.~\ref{fig:sep_1to5}) the stability of such initial condition was
maintained.
Individual simulations run time will vary slightly, depending on how
fast the secondary BH sinks toward the centre and forms a binary with
the primary (see Fig.~\ref{fig:sep_1to5}). After this happens we let
the simulations continue for a few ($\approx 5$) dynamical times to
let the system relax. Only then we analyse the results.

We designed and ran a set of simulations wide enough to probe a
variety of merging scenarios with different mass ratios and orbital
inclinations (see Tab.~\ref{tab:sims}). 
In these runs we vary the mass ratio between the primary and secondary
SMBH from 1:1 to 1:10 to explore the cosmologically relevant parameter
space. This argument assumes a constant mass ratio between the SMBHs
and the host galaxies \citep{Har04}, so that a 1:1 ratio in SMBH
masses simulates a 1:1 galaxy merger. Even if this is the case at the
beginning of a merger, the SMBHs can accrete a significant amount of
gas, changing their initial mass ratio. Since the smaller secondary
galaxy is more perturbed by the merger and suffers a higher gas inflow
toward its centre, the secondary can accrete more \citep[relatively to
  its initial mass,][]{Cal11, VWa12}. This results in the formation of
more equal mass SMBH pairs when they reach the nucleus of the merger
remnant.
For these reasons we consider our 1:10 encounters as providing a lower
limit to the actual SMBH mass ratio in such small mergers.
There is no consensus on what constitutes a major merger, which we
have represented by a 1:1 mass ratio. Different authors set limits as
high as 1:2 \citep[e.g.,][]{Bro12} or as low as 1:4
\citep[e.g.,][]{Lop12,Bun09,Mal06}. Therefore we chose to simulate a
large minor merger, just below this limit, with a mass ratio of 1:5
and a smaller case of 1:10. For even more unequal mergers, works from
\citet{Cal11,Cal09} suggest that such encounters do not lead to the
formation of a close pair of SMBHs, since the tidal field of the
primary quickly strips the secondary of most of its mass, thus
decreasing the efficiency of the dynamical friction at the early
stages of the merger and leaving the secondary wandering at kpc
separations.

\section{Simulation analysis}
\label{sec:simanalysis}

The simulations follow in time steps the positions and velocities of
the black holes and of the disc and bulge stars. Once the system has
relaxed and the distance between the primary and secondary black holes
has stabilised itself we proceed to check whether the merger remnant
still shows some photometric and kinematic signature of a nuclear
disc. For this we will produce maps for the surface density and
velocity field of the merger remnant, and analyse such maps using
\textsc{kinemetry} (Krajnovic et al. 2006) in order to detect in particular
discy isophotes and regular disc-like rotation. To help our
interpretation of any photometric or kinematic detection or
non-detection, we will also produce and visually inspect maps for the
surface density of the disc particles only at the end of our
simulations.

\placefigtwo
\placefigthree

\subsection{Stellar surface brightness and velocity maps}

The simulation outputs were first binned in regular grids to create
the equivalent to surface brightness and velocity maps. This is a
requirement of the code we used to detect the presence of discs, as it
is of others such as \textsc{IRAF} task \textit{ellipse}
\citep{Jed87}. Such grids also allow us to visually follow the merging
event and have an equivalent to an image as it would be the case in
real observations. For simplicity, the grids were made only for the
central regions where the disc was initially present, measuring 400 x
400 pc, using a grid spacing of 5 pc and a Gaussian spatial smoothing
of the same size. We produced grids for both the entire remnant and
for the disc particles only, even though the latter maps will only
serve for a qualitative assessment of the disc disruption.
Because, as we will discuss later, the viewing angle $i$ affects our
ability to detect a disc amidst the bulge light, we generated maps for
$\cos(i) = 0.0,$ 0.1, 0.2, 0.3, 0.4 and 0.5, spanning from an edge-on
view to 60$^{\circ}$. There is little or no point to analyse
projections with values of $\cos(i)$ larger than 0.6 since
\citet{Rix90} have shown that at this point it becomes very hard to
photometrically detect the presence of discs.
As a first example of these maps Fig.~\ref{fig:all_i} shows the disc,
bulge and total surface density maps with corresponding velocity maps,
for the initial conditions of our simulations.

\placefigfour
\subsection{Kinemetry}

A visual inspection of the maps for the surface brightness and
velocity fields of the merging results allows already to appreciate
the effects of the different merger simulations, but to better study
the merger remnant and quantify whether a disc would still be
detectable in it we have used \textsc{kinemetry} \citep{Kra06} to
analyse the moments of the line-of-sight velocity distribution
(LOSVD). The gridding described in the previous section was done not
only to aid visualisation and qualitative assessment but also as a
necessity for the photometric side of \textsc{kinemetry}.

The \textsc{kinemetry} algorithm works under the assumption that the
even moments of the LOSVD, such as the surface brightness, are
described by a constant profile along ellipses, and that the odd
moments, such as mean velocity, can be described by a cosine law along
such contours. In particular, we will be looking for peaks in the
coefficient of the fourth cosine term on the Fourier series expansion
of the deviations from the elliptical isophotes ($a_{4}$) in the
photometry, and for very small values of the \textit{k5/k1} ratio
(analogous to the $a_{4}$) in the velocity fields. Large $a_{4}$
values will indicate discy isophotes, whereas small \textit{k5/k1}
values will ensure the presence of a regular, disc-like rotation. As
done in other photometric tools such as the \textsc{IRAF} task ellipse
\citep{Jed87}, \textsc{kinemetry} begins by identifying the
best-fitting ellipses to the isophotes to retrieve the radial profiles
for the average surface brightness, position angle (PA), flattening
($q$) and higher moments such as the $a_{4}$. Similarly, from the odd
moments, we obtain a velocity curve (traced by the $k1$ parameter), a
kinematic PA and flattening, and the \textit{k5/k1} values
\citep[see][for more details]{Kra06}.

In Fig.~\ref{fig:d_f} we show the main outputs from
\textsc{kinemetry} for the same initial conditions shown in
Fig.~\ref{fig:all_i}. As expected, we can see $a_{4}$ and velocity
peaks in both the disc and total surface brightness (blue and black
lines respectively) whereas the bulge (red line) stays flat. On the
other hand, the \textit{k5/k1} values for the disc and total velocity
field present very low values, tracing disc-like rotation, whereas the
bulge displays a noisy \textit{k5/k1} profile. 
These are the quantities that will allow us to tell, once compared to
appropriate detection thresholds (Sec. 3.3) if the photometric and
kinematic signatures of a disc are present or not when looking at
the final stages of the interaction.

Fig.~\ref{fig:kinemetry} shows examples of the surface brightness and
velocity maps with corresponding \textsc{kinemetry} outputs for the
final stages of the 1:1, 1:5 and 1:10 mass ratios after a $i_{\rm BH}
= 30^\circ$ encounter and seen at a $60^\circ$ inclination, similar to
what was shown for the initial conditions. The bulge results have been
omitted for simplicity as they are not key to our goals. Similarly, we
do not follow the behaviour of the disc alone (blue lines) since,
although interesting, what observations trace is the surface
brightness and average velocity of all the stars. For the purpose of
detecting the signature of a disc, we will therefore look at and
analyse the surface density and kinematics of all the particles in our
simulations (black lines), in particular, the region where the disc
was originally present, which corresponds to the solid line in the
plots and is delimited by the vertical dashed line, placed at the 100
pc radius.

\placefigfive

\subsection{Disc kinemetric signatures}
\label{subsec:discsignature}

In order to assess at the end of which kind of merger events we could
still find the photometric or kinematic signatures of a disc, we
considered the results of the \textsc{kinemetry} analysis inside the
central 100 pc region originally occupied by the disc, although for
edge-on projections this radius was decreased to 50 pc when
considering the velocity $V$ and the $k5/k1$ parameter, due to
convergence problems with the \textsc{kinemetry} procedure. In
Fig.~\ref{fig:kinemetry} such a limit is indicated by a vertical
dashed line, and the \textsc{kinemetry} descriptions inside this
radius are plotted as solid lines.

To identify the presence of a disc, we will use the mean values for
the \textsc{kinemetry} parameters in such central regions, in
particular to isolate fluctuations caused only by a fraction of the
disc particles which had acquired extreme behaviours, such as those
ejected at large radii. 
Ideally, when computing such average values one would first guess the
size of the disc by locating a clear peak of the $a_4$ parameter,
which indeed is a good gauge of the disc extent (see Fig.~5 in
Paper~I). This approach would not be applicable to all our merger
remnants, however, in particular for the most conspicuos 1:1 mergers
where the $a_4$ signal is weak and noisy. On the other hand, when a
disc signature is detected, we always found an $a_4$ peak close to the
original edge at 100 pc of our initial disc, since in fact when the
disc particles remain in a disc-like configuration, the extent of such
disc remnants is similar to that of our initial conditions. This can
be appreciated in Fig.~\ref{fig:discmaps}, where we collected the
surface density distribution of the disc particles at the end of all
our simulations as seen both edge-on and from the lowest inclination
that we considered, of 60$^{\circ}$.

In the next section we will compare the average values for the $a_{4}$
parameter, the rotation velocity $V$ and for the \textit{k5/k1} ratio,
with corresponding threshold values that indicate the presence of a
disc. For the $a_{4}$ parameter we considered a generous minimum
average value of 0.02 for detecting discy isophotes in our
simulations, which is comparable to what is observed in real
NSD-hosting galaxies \citep[see e.g.,][]{Mor04, Piz02}. As regards
kinematics, we require an average velocity above $10\rm km s^{-1}$ to
detect global rotation, in line with typical errors on this quantity,
and average $k5/k1$ values below 0.04 to recognise the presence of
disc-like motions. The latter threshold, which lies just above the
typical average value found in our initial conditions, is taken from
\citet{Kra11} who based their judgement on integral-field data for 260
early-type galaxies from the ATLAS$^{\rm 3D}$ survey \citep{Cap11}.

Fig.~\ref{fig:kinresults} presents the central average values for the
$a_{4}$, $V$ and \textit{k5/k1} parameters measured in the
surface-density and velocity maps for the final step of each of our
simulations as observed from different viewing angles from edge-on
down to a 60$^{\circ}$ inclination. The different panels in these
figures group together simulations with similar impact angle $i_{\rm
  BH}$. In each panel the size of the circles show the average
$a_{4}$, $V$ and \textit{k5/k1} values, where the size of these
symbols correspond to the mass-ratios of the merger events (i.e., with
largest circles showing 1:1 mergers). Finally, the asterisks in
Fig.~\ref{fig:kinresults} also show the average values of these
parameters from our initial conditions. These are the figures that
will be discussed in the next section.

\section{Results}
\label{sec:results}

\placefigsix

Photometrically, the search for the presence of nuclear discs
generally starts with the visual inspection of unsharped images or
structure maps, which highlight small scale features
\citep[e.g.,][Paper~I]{Piz02}. This is then followed by a more
quantitive analysis of the shape of the stellar isophotes, looking in
particular for positive peaks in the $a_4$ parameter. With
integral-field spectroscopic data it will further be possible to look
for rotating structures that may not always show up photometrically,
which are likely to be discs when finding very low values for the
$k5/k1$ kinemetric parameter that would indicate a regular, disc-like
kinematics.

\subsection{Visual inspection of disc remnants}

Before analysing our kinemetric results on the global outcome of our
simulations, we produce maps for the surface density of the disc
particles only and inspect such disc remnants. This we do in order not
only to visually assess the disruption of the disc but most important
to gain some insights on our kinemetric findings and the detection of
disc signatures.
As one would expect, Fig.~\ref{fig:discmaps} shows that major mergers
(with a 1:1 mass ratio, left) have the most dramatic impact on the
disc, as its particles get scattered at large radii and vertically
leading to lower surface density values and thicker remnants compared
to intermediate and minor mergers (with 1:5 and 1:10 mass ratio,
center and right). Fig.~\ref{fig:discmaps} further indicates that
prograde encounters (top panels) affect the disc structure more than
their retrograde counterparts (lower panels), which can be noticed by
looking in particular at the 1:5 merger disc remnants.
This visual inspection also reveals interesting structures in the disc
remnants, such as circumnuclear rings or the accumulation of stars in
central clusters.

\subsection{Incidence of disc kinemetric signatures}

Having defined earlier in Sec.~\ref{subsec:discsignature} the
thresholds for the detection of a disc in our simulations based on
average values for the $a_4$, $V$ and $k5/k1$ parameters, and after
checking that all our simulations reached a post-interaction relaxed
state, we now proceed to assess in which kind of simulation and down
to which inclination we can still find clear photometric and/or
kinematic signatures of a disc.

Starting from the plots for the $a_4$ parameter (top panels in
Fig.~\ref{fig:kinresults}), we re-observe the trend whereby the
photometric signature of a disc gets more easily erased during the
interactions involving a more massive secondary black hole. Indeed,
1:1 mergers leave little or no trace of a nuclear disc in 81$\%$ of
the cases whereas the final steps of the 1:5 and 1:10 mergers display
average $a_4$ values above the threshold suggestive of a disc
structure in 73$\%$ and 86$\%$ of the cases. Prograde encounters would
appear to be generally more destructive, as can be noticed by
comparing the trends for the 1:5 and 1:10 remnants for $i_{\rm BH}=30$
and 60$^{\circ}$ with their counterparts at $i_{\rm BH}$ values of
120$^{\circ}$ and 150$^{\circ}$, or by observing that most of the 1:1
merger remnants showing discy isophotes are found in the case of a
perfectly retrograde encounter ($i_{\rm BH}=180$). In fact, the
edge-on projection of the 1:1 prograde merger ($i_{\rm BH}=0$) owes
its large $a_4$ values only to a distinct ring-like structure in the
disc remnant (Fig.~\ref{fig:discmaps}), which would not be mistaken
for a disc upon a simple visual inspection.

The difference between prograde and retrograde can be understood
considering that stars are more easily deflected by the secondary
black hole when this travels alongside rather than against them, which
is also why dynamical friction on the secondary is more efficient in
the prograde cases making it sink faster towards the centre. This can
be appreciated back in Fig.~\ref{fig:sep_1to5} which shows the
evolution of the black holes separation as a function of time and
where the prograde tracks for $i_{\rm BH} = 0$, 30 and 60 indicate
always a faster sinking than their retrograde counterparts for $i_{\rm
  BH} = 180$, 150 and 120. \citet{Mei13} provide another example of
this behaviour in the somehow different context of a binary black-hole
interaction within a non-rotating bulge, which ends up showing a
central rotating structure as the coalescing black holes evacuated
preferentially the stars that co-rotate with them. The larger impact
of a prograde encounter on discs was also already shown by
\citet{Vel99} in the case of Milky Way-like galactic discs and
satellites.

Moving on to the trends for the peak velocity $V$ traced by our
kinemetric fit (middle panels in Fig.~\ref{fig:kinresults}), we
observe that - except for the planar and polar 1:1 encounter and the
planar prograde 1:5 merger - some level of bulk stellar rotation would
always remain detectable in the remnant, even in cases where no
photometric signature of the disc was identified in the $a_4$
profiles. This suggests, as one would expect, that while perturbing
the radial and vertical structure of a thin disc is relatively easy,
entirely erasing the signature of the rotation of its stars is much
more difficult, and requires a rather dramatic scattering of the disc
stars at larger radii. This is for instance the case of the
aforementioned ring-like structure in the 1:1 planar prograde merger,
which is all that is left of the disc and does not contain a
sufficient number of particles to imprint a signature of bulk rotation
in the merger remnant, even when seen edge-on.

Not all rotating remnants show regular disc-like kinematics,
however. Looking finally at the average values for the $k5/k1$
parameters (lower panels in Fig.~\ref{fig:kinresults}) shows in
particular that nearly half of the 1:1 remnants do not rotate
regularly when they do show bulk rotation. On the other hand, except
for a few projections of the prograde encounter, the remnants of 1:5
interactions generally display regular kinematics, with this being
always the case of the 1:10 remnants.

Overall, a sufficiently ample and regular rotation is found in 38$\%$,
72$\%$ and all of the remnants of the 1:1, 1:5 and 1:10 mergers,
respectively. 
Many of these also show photometric disc signature that would suggest
the presence of a rather thin disc component, whereas conversely the
absence of a photometric signature, which occurs in particular in 1:1
mergers, could lead to infer the presence of a rather thick disc (as
Fig.~\ref{fig:kinresults} indeed shows).
Based on this limited number of simulations we can draw an initial
general picture for the chance of nuclear stellar discs surviving a
galaxy merger and of being detected, considering in particular that
co-planar encounters are, for a purely probabilistic perspective, much
less likely to happen than merger events where the secondary black
hole initially moved in an highly inclined orbit. We can account for
this effect by correcting for a sin$(i_{\mathrm{BH}}$) factor our raw
fractions of remnants with kinematic and photometric disc signatures,
which effectively removes the perfectly co-planar cases and
accentuates the impact of polar or nearly polar impacts. Since the
former simulations tend to be the more disruptive for the disc, this
correction tends to produce larger detection rates for discy or
regularly rotating structures. These fractions do not vary
dramatically, however, in particular since the perfectly polar impact
can also be quite damaging for the disc.
The values for both our raw and sin$(i_{\mathrm{BH}}$)-corrected
fractions of merger remnants with photometric discy features or
regular velocity fields are reported in Tab.~\ref{tab:frac}.

\placetabtwo

\section{Discussion}
\label{sec:conclusions}

Using a suite of numerical simulations we have explored the fragility
of nuclear stellar discs in order to establish their usefulness as
tools to constrain merging histories of early-type galaxies, looking
in particular at the same photometric and kinematic signatures that
observers would regard as indicative of the presence of a NSD.
Our results show that NSDs are destroyed in major merger situations
whereas photometric and, even more generally, kinematic signatures of
a disc are still detectable in the remnants of minor mergers (see
Tab.~\ref{tab:frac}). Integral-field studies should therefore prove
particularly useful to compile a more complete census of NSDs than
already done based solely on HST images in Paper~I. In particular, new
instruments such as MUSE \citep{Bac10}, with its superior sensitivity
and the ability to work in conjunction with adaptive optics, will be
very useful for detecting NSDs. Additionally, MUSE should also prove
an excellent tool for dating the age of these structures, improving on
the first experiment of Sarzi et al. (in preparation) based on VIMOS
data.

A large sample of NSDs with well constrained values for their stellar
age, combined with the knowledge gained here on the fragility of such
structures, should allow to paint a more detailed picture for the
assembly history of galaxies.
For instance, finding evidence for a NSD both kinematically and
photometrically - that is - finding the signature of a thin disc,
would suggest according to our analysis that there is only a
$\sim20\%$ chance that the host galaxy of such a disc would have
experienced a major merger since the formation epoch of the disc, as
indicated by its stellar age. In fact, our estimated incidence of
nuclear-disc hosting galaxies that would still show evidence for a
disc after a merger are most likely conservative limits, in particular
since we have not considered here the possible presence of a stellar
cusp around the secondary incoming SMBH, and the additional negative
impact that would come with such an added stellar mass.

Galaxy evolution models predict merging trees of a combination of
mergers \citep[e.g.,][]{Kho06a, Kho06b, Gon11} and there is a great
debate on the influence of major and minor mergers in the size and
mass growth of galaxies, with current theory favouring 1:5 - 1:10
mergers \citep[e.g.,][]{Ose12,Bed13}. This corresponds to the minor
merger range of mass ratios studied in this work.  Yet major mergers
cannot be discarded, as other studies \citep[e.g.,][]{Lid13} have
suggested that in recent epochs they may have caused large mass
increases. Therefore, the ability of these NSDs to discriminate
between merger scenarios is central to today's questions.

We are aware that the set up of the simulations presented here is
simplified. For instance considering a Mestel profile for the nuclear
disc and a purely spherical Plummer distribution for the bulge does
not convey the complexity of real nuclei where e.g., the disc is
better described by an exponential law and bulges can be flattened and
present both core and cuspy profiles. Yet, with a bulge mass
approximately 5 times that of the disc within 100 pc, the
corresponding disc to total surface brightness ratio of our initial
model does span the observed values between 20 and 40\%
\citep[e.g.,][]{Mor04}, depending on the inclination. We therefore
regard this work a first exploratory study, which could be improved on
the one hand by exploring a larger parameter space, for instance by
varying the mass of the nuclear disc compared to its host bulge and
central SMBH, or on the other hand by considering a
more realistic model with a stellar cusp around the secondary black
hole. The actual amount of stellar mass bound to the sinking nucleus
of a satellite galaxies may well remain a free parameter in a future
analysis, unless new insights will emerge from larger sets of
simulations such as those of \citet{Cal09}, but in general we expect
that the presence of a stellar cusp around the secondary will further
contribute to the disc disruption.

To conclude, we note that better simulations may as well shed more
light also on the origin of the kinematically decoupled cores (KDC)
that are often found in early-type galaxies \citep[e.g.,][]{Ben88,
  Fra89, McD06}. Indeed a significant number of our simulated
encounters result in remnants that display regular rotation,
characterised even by a low average $k5/k1$ value, despite not showing
any photometric $a_4$ signatures of discy structures. 
This resembles the situation observed for the KDCs of massive
elliptical galaxies \citep[see, e.g., Fig~C4 of][]{Kra11}, where
indeed a distinct kinematic central pattern is never accompanied by a
significant change in the isophotes orientation, flattening or shape
in the same regions. To us, this similarity suggests a possible link
between KDCs and puffed-up thick discs, and a closer look to the
higher-order moments of the stellar LOSVD, both in future simulations
and in high-quality integral-field data, should provide more valuable
insights on the nature of these structures.

\section*{Acknowledgements}
We are grateful to the anonymous referee for carefully reviewing our
manuscript, which considerably helped in improving our work. We are
also indebted to R.~L.~D. Davies for also providing useful comments.


\end{document}